\documentstyle[11pt,newpasp,twoside,epsf]{article}
\def\edcomment#1{\iffalse\marginpar{\raggedright\sl#1\/}\else\relax\fi}
\reversemarginpar

\begin{document}
\title{Star Cluster Formation in Extreme Starburst Environments}
 \author{Richard de Grijs}
\affil{Department of Physics \& Astronomy, The University of Sheffield, 
Hicks Building, Hounsfield Road, Sheffield S3 7RH, UK}

\begin{abstract}
The currently available empirical evidence on the star formation processes in
the extreme, high-pressure environments induced by galaxy encounters, mostly
based on high-resolution {\sl Hubble Space Telescope} imaging observations,
strongly suggests that star {\it cluster} formation is an important and
perhaps even the dominant mode of star formation in the starburst events
associated with galaxy interactions.
\end{abstract}

\section{Gravitational Interaction Induced Star (Cluster) Formation}

Close encounters between gas-rich galaxies often have devastating effects on
the state of their interstellar medium (ISM), in particular of the ISM of
either the smaller or of the most gas-rich of the interacting galaxies (e.g.,
de Grijs et al. 2001, 2003). The ram pressure caused by the time-varying
gravitational potential induced by the interaction causes free-floating giant
molecular clouds in the affected galaxy's ISM to collapse and enter a phase of
violent star formation. This process is particularly well illustrated by the
ongoing interaction between NGC 6745 and its small northern companion galaxy,
``NGC 6745c'' (cf. de Grijs et al. 2003), where dynamical and kinematic
evidence from H{\sc i} observations obtained with the Very Large Array support
the scenario that the small companion galaxy has traveled from the south east
across the eastern edge of the main galaxy to its current location north of
the main galaxy, in its wake leaving a frenzy of very violent and active star,
and in particular star {\it cluster}, formation.

In de Grijs et al. (2003b) we used the ages, masses and metallicities of the
rich young star cluster system in NGC 6745, based on the analysis of
multi-passband archival {\sl Hubble Space Telescope (HST)} observations, to
derive its cluster formation history and subsequent evolution. We derive a
median age, and thus an estimate of the starburst duration, of $\sim 10$ Myr.
NGC 6745 contains a significant population of high-mass ``super star clusters
(SSCs)'', with masses in the range $6.5 \la \log( M_{\rm cl}/{\rm M}_\odot )
\la 8.0$. We caution, however, that these massive SSC candidates may not be
gravitationally bound objects, but more diffuse star forming regions or
aggregates of multiple unresolved clusters instead. Nevertheless, we measure
an effective radius for the most massive object ($M_{\rm cl} \simeq 5.9 \times
10^8 {\rm M}_\odot$) of only $R_{\rm eff} \sim 16$ pc. However, this object
appears very elongated, or may in fact be a double cluster. We should keep in
mind, of course, that this high mass estimate is a strong function of the
(low) metallicity assumed; if we had assumed solar metallicity for this
object, the derived age would have been significantly smaller ($\sim 10-20$
Myr vs. $\sim 1$ Gyr), and the mass could be smaller by a factor of $\ga 10$.
Even so, if we could confirm this mass estimate spectroscopically, either of
the subcomponents would be the most massive cluster known to date,
significantly exceeding cluster W3 in NGC 7252, which has a mass of about
$(3-18) \times 10^7 {\rm M}_\odot$, depending on the age, metallicity and IMF
assumed (Schweizer \& Seitzer 1998; Maraston et al. 2001).

The suggestion that such massive objects (either star clusters, or extended
star-forming regions) form preferentially in the extreme environments of
interacting galaxies, is supported quantitatively by our analysis (de Grijs et
al. 2003a) of pixel-by-pixel colour-magnitude and colour-colour diagrams of
the Mice and Tadpole interacting galaxies (NGC 4676 and UGC 10214,
respectively), based on a subset of the (archival) {\sl HST} Advanced Camera
for Surveys Early Release Observations, which provide a powerful technique to
explore and deduce the star and star cluster formation histories of galaxies
at moderate distances. In each interacting system we found some 40 bright
young star clusters, with a characteristic mass of $\sim 3 \times 10^6 {\rm
M}_\odot$, which are spatially coincident with blue regions of active star
formation in their tidal tails and spiral arms. We showed that star cluster
formation is a major mode of star formation in galaxy interactions, with $\ga
35$\% of the active star formation in encounters occurring in star clusters.
In particular, the tidal tail of the Tadpole system is dominated by blue
star-forming regions, which occupy some 60\% of the total area covered by the
tail and contribute $\sim 70$\% of the total flux in the F475W filter
(decreasing to $\sim 40$\% in F814W).

Finally, in the nearby (ultra)luminous infrared galaxy NGC 6240, we detect a
population of massive young star clusters, with tentative evidence (based on
statistical considerations) that the more massive clusters are found closer to
the galaxy's double (or perhaps triple) nucleus and the most intense starburst
region (Pasquali, de Grijs, \& Gallagher 2003).

Thus, the currently available empirical evidence, mostly based on
high-resolution {\sl HST} imaging observations, strongly suggests that star
cluster formation is an important and perhaps even the dominant mode of star
formation in the starburst events associated with galaxy interactions.

\acknowledgements I acknowledge invaluable collaborations with Peter Anders,
Nate Bastian, Uta Fritze-v. Alvensleben, Jay Gallagher, Henny Lamers, and Anna
Pasquali.

\newpage

\title{``Super'' Star Clusters}
 \author{Richard de Grijs}
\affil{Department of Physics \& Astronomy, The University of Sheffield, 
Hicks Building, Hounsfield Road, Sheffield S3 7RH, UK}

\begin{abstract}
The production of ``super star clusters'' (SSCs; luminous, compact star
clusters) seems to be a hallmark of intense star formation, particularly in
interacting and starburst galaxies. Their sizes, luminosities, and mass
estimates are entirely consistent with what is expected for young Milky
Way-type globular clusters (GCs). SSCs are important because of what they can
tell us about GC formation and evolution (e.g., initial characteristics and
early survival rates). They are also of prime importance as probes of the
formation and (chemical) evolution of their host galaxies, and of the initial
mass function in the extreme environments required for cluster formation.
Recent evidence lends support to the scenario that Milky Way-type GCs
(although more metal rich), which were once thought to be the oldest building
blocks of galaxies, are still forming today.
\end{abstract}

\section{``Super'' or normal?}

One of the main contributions to date of the {\sl Hubble Space Telescope
(HST)} to the field of stellar populations in nearby galaxies has been the
discovery of numerous dense stellar objects resembling star clusters with
properties similar to those predicted for the progenitors of the old globular
cluster (GC) population in the Milky Way and other nearby galaxies.

These objects are often confusingly referred to as ``super star clusters''
(SSCs), by virtue of their high luminosities and compact sizes. They have been
found in a wide variety of galactic environments, ranging from quiescent dwarf
and irregular or amorphous galaxies to large, gas-rich spiral galaxies
involved in large-scale gravitational interactions and mergers, and in the
starburst events associated with them (see de Grijs, O'Connell, \& Gallagher
2001 for an overview). However, the question arises of whether these objects
are indeed ``super'' star clusters, in terms of either their integrated
luminosity or their total mass. If they are indeed the progenitors of Milky
Way-type GCs, assuming that they have the potential to survive for a Hubble
time, then their high luminosities at their correspondingly young ages (of up
to $\sim 1$ Gyr, in general) are simply conform the expectations of any modern
flavour of simple stellar population theory.

Indeed, the mass distributions of most of these young star cluster populations
do not extend significantly beyond that of the Milky Way GC population (which
is generally used as a benchmark), with a few exceptions (e.g., NGC 7252-W3:
Schweizer \& Seitzer 1998, Maraston et al. 2001; NGC 6745: de Grijs et al.
2003c; some of the Antennae clusters, e.g., Mengel et al. 2002), although
their existence might simply be due to stochastic effects. Nevertheless, the
latter objects may therefore truly be super{\it massive} objects. With the
exception of these few clusters, one should perhaps not consider the overall
mass distribution of a given cluster population compared to that of the almost
universal GC mass function in a wide variety of galaxies hosting such old
objects, but instead consider these remarkable clusters in the context of
their own parent population. This exercise leads us to realise that in a
number of (predominantly) dwarf and irregular galaxies the overall cluster
population is host to a few clusters that are significantly more massive than
any of the other clusters (although not necessarily more massive than the
high-mass wing of the Milky Way GC system), such as observed in NGC 1705 (NGC
1705-I, Ho \& Filippenko 1996), NGC 1569 (SSCs A and B; see, e.g., Hunter et
al. 2000, and Anders et al 2003 for comparative analyses in the context of the
galaxy's overall cluster population), and M82 (M82-F; e.g., Smith \& Gallagher
2001). Therefore, the assignation ``super'' appears to be merely a relative
qualification.

In view of the confusing nomenclature, I will henceforth refer to these
objects as {\sl Young Massive Star Clusters (YMCs)}.

\section{Survival to old age?}

Although YMC populations are often assumed to be GC-type progenitors, their
survival for a Hubble time is by no means guaranteed. In fact, this depends
crucially on the slope of the stellar initial mass function (IMF) governing
these clusters. High-resolution spectroscopy can -- for the nearest YMC
systems -- be utilised to derive dynamical mass estimates and, combined with
integrated luminosity measurements from e.g. {\sl HST} imaging, one can derive
mass-to-light (M/L) ratios for a small number of clusters at a time. In their
comparison of the M/L ratios at the corresponding ages for a handful of the
brightest YMCs, Smith \& Gallagher (2001) and Mengel et al. (2002) showed
convincingly that a number of them appear to have IMF slopes that are
significantly too shallow for the clusters to survive for any longer than
roughly the next Gyr. Thus, these objects are unlikely to become GC analogues.

Instead of going through the cumbersome exercise of measuring individual M/L
ratios, one can approach this problem statistically, by analysing the
potential of a given cluster population to survive for a Hubble time. The
currently most popular models for the dynamical evolution of star clusters
predict that the power-law Cluster Luminosity Functions (CLFs) characteristic
of YMC systems will be transformed rapidly into the universal Gaussian CLFs of
old Milky Way-type GC systems. In a recent paper (de Grijs, Bastian, \& Lamers
2003a; see also de Grijs, Bastian, \& Lamers 2003b), we provided the first
evidence for a turn-over in the intermediate-age, approximately 1 Gyr-old CLF
in the centre of the nearby starburst galaxy M82, which very closely matches
the universal CLFs of old Milky Way-type GC systems. This is likely to remain
virtually unchanged for a Hubble time. We also showed that with the very short
characteristic cluster disruption time-scale governing the centre of M82 (de
Grijs et al. 2003b), its cluster mass distribution will evolve towards a
higher characteristic mass scale than for the Galactic GCs by the time it
reaches a similar age. We argue, therefore, that this evidence, combined with
the similar cluster sizes (de Grijs et al. 2001), lends strong support to a
scenario in which the current central M82 cluster population will eventually
evolve into a significantly depleted old Milky Way-type GC system dominated by
a small number of high-mass clusters. This implies that GC progenitors, which
were once thought to be the oldest building blocks of galaxies, are still
forming today in galaxy interactions and mergers. However, they will likely be
more metal-rich than the present-day old GC systems. This connection between
young or intermediate-age star cluster systems and old GCs lends strong
support to the hierarchical galaxy formation scenario.

M82's proximity, its shortest known cluster disruption time-scale of any
galaxy, and its well-defined peak of cluster formation make it an ideal
candidate to probe the evolution of its star cluster system to fainter
luminosities, and thus lower masses, than has been possible for any galaxy
before.

\acknowledgements I acknowledge invaluable collaborations with Peter Anders,
Nate Bastian, Uta Fritze--v. Alvensleben, Jay Gallagher, and Henny Lamers.

\end{document}